\documentclass[12pt]{article}
\usepackage{color}
\begin{document}
\baselineskip=24pt
\hoffset = -0.3truecm
\voffset = -1.1truecm

\title{\bf \textcolor{blue}{
MONOPOLE-ANTIMONOPOLE AND VORTEX RINGS}}

\author{
{\bf Rosy Teh and Khai-Ming Wong}\\
{\normalsize School of Physics, Universiti Sains Malaysia}\\
{\normalsize 11800 USM Penang, Malaysia}\\
{\normalsize e-mail: rosyteh@usm.my}}

\date{March 2005}
\maketitle

\begin{abstract}
The SU(2) Yang-Mills-Higgs theory supports the existence of monopoles, antimonopoles, and vortex rings. In this paper, we would like to present new exact static antimonopole-monopole-antimonopole (A-M-A) configurations. The net magnetic charge of these configurations is always negative one, whilst the net magnetic charge at the origin is always positive one for all positive integer values of the solution's parameter $m$. However, when $m$ increases beyond one, vortex rings appear coexisting with these A-M-A configurations. The number of vortex rings increases proportionally with the value of $m$. They are located in space where the Higgs field vanishes along rings. We also show that a single point singularity in the Higgs field does not necessarily corresponds to a structureless 1-monopole at the origin but to a zero size monopole-antimonopole-monopole (\textcolor{blue}{MAM}) structure when the solution's parameter $m$ is odd. This monopole is the Wu-Yang type monopole and it possesses the Dirac string potential in the Abelian gauge. 
These exact solutions are a different kind of BPS solutions as they satisfy the first order Bogomol'nyi equation but possess infinite energy due to a point singularity at the origin of the coordinate axes. They are all axially symmetrical about the z-axis. 
\end{abstract}


\section{Introduction}
\label{sect:1}
The SU(2) Yang-Mills-Higgs (YMH) field theory, with the Higgs field in the adjoint representation possesses magnetic monopole, multimonopole, antimonopoles, and vortex rings solutions \cite{kn:1}-\cite{kn:8}. The only spherically symmetric monopole solution is the unit magnetic charge monopole. The 't Hooft-Polyakov monopole solution with non zero Higgs mass and Higgs self-interaction is numerically, spherically symmetrical \cite{kn:1}-\cite{kn:2}. Multimonopole solutions possess at most axial symmetry \cite{kn:3}. 

The model with non vanishing Higgs vacuum expectation value but vanishing Higgs potential possesses exact monopole and multimonopole solutions which can be obtained by solving the first order Bogomol'nyi equations \cite{kn:11}. These solutions satisfying the Bogomol'nyi-Prasad-Sommerfield (BPS) limit possess minimal energies. Exact monopole and multimonopoles solutions exist in the BPS limit \cite{kn:2} - \cite{kn:3} whilst outside the BPS limit, when the Higgs field potential is non vanishing only numerical solutions are known.
Asymmetric multimonopole solutions with no rotational symmetry  are numerical solutions even in the BPS limit \cite{kn:4}.

At present, the different exact configurations of monopoles found are the BPS multimonopole solutions of magnetic charges $M$ greater than unity with all the magnetic charges superimposed into a single point in space \cite{kn:3}. These superimposed multimonopole solutions possess axial and mirror symmetries. 
Following these works, finitely separated 1-monopoles were also constructed \cite{kn:4}-\cite{kn:5}.

Numerical axially symmetric non-Bogomol'nyi monopole-antimonopole chain solutions were also found to exist both in the limit of a vanishing Higgs potential as well as in the presence of a finite Higgs potential. Recently, numerical BPS axially symmetric vortex rings solutions have also been reported \cite{kn:6}.

We have reported on the existence of a different type of BPS static monopole-antimonopole solution in Ref.\cite{kn:7}. This solution which is exact and axially symmetric, represents two separate antimonopoles located at equal distances along the z-axis from a 1-monopole which is located at the origin.  We have also shown that the extended ansatz of Ref.\cite{kn:7} possesses more multimonopole-antimonopole configurations, together with their anti-configurations \cite{kn:8}. These configurations possess either radial, axial, or mirror symmetries about the z-axis and they represent different combinations of monopoles, multimonopole, and antimonopoles. 

In general, configurations of the YMH field theory with a unit magnetic charge are spherically symmetric \cite{kn:1}, \cite{kn:2}, whilst multimonopole configurations with magnetic charges greater than unity possess at most axial symmetry \cite{kn:3}. However we have emphasized in a recent work \cite{kn:9} that unit magnetic charge configurations are not necessarily spherically symmetric. By employing the ansatz of Ref.\cite{kn:7} we have found exact unit magnetic charge solution that does not even possess axial symmetry but only mirror symmetry about a vertical plane through the z-axis.  However the converse is true and it has been shown that multimonopole solutions cannot possess spherical symmetry \cite{kn:10}. We would also like to mention that within the ansatz of Ref.\cite{kn:7}, half-monopole solutions have also been reported \cite{kn:9}.

In this paper we would like to present new static axially symmetric antimonopole-monopole-antimonopole (A-M-A) configurations of the SU(2) YMH theory with the Higgs field in the adjoint representation. Here the Higgs field vanishes either at points corresponding to antimonopoles or at rings corresponding to vortex loops.   
The net magnetic charge of these configurations is always negative one, whilst the net magnetic charge at the origin, $r=0$, is always positive one for all positive integer values of the solution's parameter $m$. However, when $m$ increases beyond one, vortex rings appear coexisting with these A-M-A configurations. The number of vortex rings in the configuration is equal to $(m-1)$ where $m\geq 1$. They are located horizontally in space where the Higgs field is zero along rings. Hence this family of solutions all lies in the topologically non trival sector with topogical charge negative one. 

The two antimonopoles of the solutions are located at the two zeros of the Higgs field along the z-axis, whilst the Wu-Yang type 1-monopole is located at a point singularity of the Higgs field at the origin.
We also show that this single point singularity in the Higgs field need not corresponds to a structureless 1-monopole at the origin but to a zero size  monopole-antimonopole-monopole (\textcolor{blue}{MAM}) structure when $m$ is odd.
These exact solutions are a different kind of BPS solutions as they satisfy the first order Bogomol'nyi equation but possess infinite energy due to a point singularity at the origin of the coordinate axes. 

We briefly review the SU(2) Yang-Mills-Higgs field theory and discussed the boundary conditions of these solutions in the section \ref{sect:2}. We discussed the magnetic ansatz and its formulation in section \ref{sect:3} and present the solutions in section \ref{sect:4}. We end with some comments in section \ref{sect:5}.

\section{The SU(2) Yang-Mills-Higgs Theory}
\label{sect:2}
The SU(2) group admits the triplet Yang-Mills gauge fields potential $A^a_\mu$ which when coupled to a scalar Higgs triplets field $\Phi^a$ in 3+1 dimensions gives the SU(2) YMH theory \cite{kn:12}. The index $a$ is the SU(2) internal space index and for a given $a$, $\Phi^a$ is a scalar whereas $A^a_\mu$ is a vector under Lorentz transformation. The SU(2) YMH Lagrangian is given by
\begin{equation}
{\cal L} = -\frac{1}{4}F^a_{\mu\nu} F^{a\mu\nu} + \frac{1}{2}D^\mu \Phi^a D_\mu \Phi^a - \frac{1}{4}\lambda(\Phi^a\Phi^a - \frac{\mu^2}{\lambda})^2, 
\label{eq.1}
\end{equation}

\noindent where the Higgs field mass, $\mu$, and the strength of the Higgs potential, $\lambda$, are constants. The vacuum expectation value of the Higgs field is then $\mu/\sqrt{\lambda}$. The Lagrangian (\ref{eq.1}) is gauge invariant under the set of independent local SU(2) transformations at each space-time point.
The covariant derivative of the Higgs field and the gauge field strength tensor are given respectively by 
\begin{eqnarray}
D_{\mu}\Phi^{a} &=& \partial_{\mu} \Phi^{a} + \epsilon^{abc} A^{b}_{\mu}\Phi^{c},\\
\label{eq.2}
F^a_{\mu\nu} &=& \partial_{\mu}A^a_\nu - \partial_{\nu}A^a_\mu + \epsilon^{abc}A^b_{\mu}A^c_\nu.
\label{eq.3}
\end{eqnarray}
Since the gauge field coupling constant $g$ can be scaled away, we can set $g$ to one without any loss of generality. The metric used is $g_{\mu\nu} = (- + + +)$. The SU(2) internal group indices $a, b, c$ run from 1 to 3 and the spatial indices are $\mu, \nu, \alpha = 0, 1, 2$, and $3$ in Minkowski space.

The equations of motion that follow from the Lagrangian (\ref{eq.1}) are
\begin{eqnarray}
D^{\mu}F^a_{\mu\nu} &=& \partial^{\mu}F^a_{\mu\nu} + \epsilon^{abc}A^{b\mu}F^c_{\mu\nu} = \epsilon^{abc}\Phi^{b}D_{\nu}\Phi^c,\nonumber\\
D^{\mu}D_{\mu}\Phi^a &=& -\lambda\Phi^a(\Phi^{b}\Phi^{b} - \frac{\mu^2}{\lambda}).
\label{eq.4}
\end{eqnarray}
The tensor to be identified with the Abelian electromagnetic field, as proposed by 't Hooft \cite{kn:1}, \cite{kn:13} is
\begin{eqnarray}
F_{\mu\nu} &=& \hat{\Phi}^a F^a_{\mu\nu} - \epsilon^{abc}\hat{\Phi}^{a}D_{\mu}\hat{\Phi}^{b}D_{\nu}\hat{\Phi}^c\nonumber\\
&=& \partial_{\mu}A_\nu - \partial_{\nu}A_\mu - \epsilon^{abc}\hat{\Phi}^{a}\partial_{\mu}\hat{\Phi}^{b}\partial_{\nu}\hat{\Phi}^c,
\label{eq.5}
\end{eqnarray}
where $A_\mu = \hat{\Phi}^{a}A^a_\mu,~~\hat{\Phi}^a = \Phi^a/|\Phi|,~~|\Phi| = \sqrt{\Phi^{a}\Phi^{a}}$. Hence the Abelian electric field is $E_i = F_{0i}$, and the Abelian magnetic field is $B_i = -\frac{1}{2}\epsilon_{ijk}F_{jk}$, where the indices, $i, j, k = 1, 2, 3$. 
The topological magnetic current, which is also the topological current density of the system is \cite{kn:13} 
\begin{equation}
k_\mu = \frac{1}{8\pi}~\epsilon_{\mu\nu\rho\sigma}~\epsilon_{abc}~\partial^{\nu}\hat{\Phi}^{a}~\partial^{\rho}\hat{\Phi}^{b}~\partial^{\sigma}\hat{\Phi}^c.
\label{eq.6}
\end{equation}
Therefore the corresponding conserved topological magnetic charge is
\begin{eqnarray}
M & = & \int d^{3}x~k_0 = \frac{1}{8\pi}\int \epsilon_{ijk}\epsilon^{abc}\partial_{i}\left(\hat{\Phi}^{a}\partial_{j}\hat{\Phi}^{b}\partial_{k}\hat{\Phi}^{c}\right)d^{3}x\nonumber\\
& = & \frac{1}{8\pi}\oint d^{2}\sigma_{i}\left(\epsilon_{ijk}\epsilon^{abc}\hat{\Phi}^{a}\partial_{j}\hat{\Phi}^{b}\partial_{k}\hat{\Phi}^{c}\right)\nonumber\\
& = & \frac{1}{4\pi} \oint d^{2}\sigma_{i}~B_i. 
\label{eq.7}
\end{eqnarray}

Our work is restricted to the static case where $A^a_0 = 0$ with massless Higgs field and vanishing self-interaction.
The magnitude of the Higgs field vanishes as $1/r$ at large $r$. However, this does not affect the Abelian magnetic field of the solutions as this magnetic field depends only on the unit vector of the Higgs field. It is in this limit that the solutions are solved using both the second order Euler-Lagrange equations and the Bogomol'nyi equations, $B^a_i \pm D_i \Phi^a = 0$. 
The $\pm$ sign corresponds to monopoles and antimonopoles respectively for the usual BPS solutions \cite{kn:13}. In our case, the A-M-A configuration is solved with the $+$ sign and its anti-configuration, that is, the M-A-M configuration is solved with the $-$ sign \cite{kn:8}. 

\section{The Ansatz and its Formulation}
\label{sect:3}
The static gauge fields and Higgs field which will lead to the axially symmetric vortex rings solutions are given respectively by \cite{kn:7}
\begin{eqnarray}
A_{\mu}^a &=& \frac{1}{r}\psi(r)\left(\hat{\theta}^{a}\hat{\phi}_\mu - \hat{\phi}^{a}\hat{\theta}_{\mu}\right)  + \frac{1}{r}R(\theta)\left(\hat{\phi}^{a}\hat{r}_{\mu} - \hat{r}^{a}\hat{\phi}_{\mu}\right),\nonumber\\
\Phi^a &=& \Phi_{1}~\hat{r}^a + \Phi_{2}~\hat{\theta}^a,
\label{eq.8}
\end{eqnarray}
\noindent where $\Phi_1 = \frac{1}{r}\psi(r), ~\Phi_2 = \frac{1}{r}R(\theta)$. The spherical coordinate orthonormal unit vectors, $\hat{r}^{a}$, $\hat{\theta}^{a}$, and $\hat{\phi}^{a}$ are defined by 
\begin{eqnarray}
\hat{r}^{a} &=& \sin\theta ~\cos \phi ~\delta^a_1 + \sin\theta ~\sin \phi ~\delta^a_2 + \cos\theta \delta^a_3,\nonumber\\
\hat{\theta}^{a} &=& \cos\theta ~\cos \phi ~\delta^a_1 + \cos\theta ~\sin \phi ~\delta^a_2 - \sin\theta ~\delta^a_3,\nonumber\\
\hat{\phi}^{a} &=& -\sin \phi ~\delta^a_1 + \cos \phi ~\delta^a_2,
\label{eq.9}
\end{eqnarray}
\noindent where $r=\sqrt{x^i x_i}$, ~$\theta=\cos^{-1}(x_3/r)$, and~$\phi=\tan^{-1}(x_2/x_1)$. The gauge field strength tensor and the covariant derivative of the Higgs field are given respectively by
\begin{eqnarray}
F^a_{\mu\nu} &=& \frac{1}{r^2}\hat{r}^{a}\left\{\dot{R} + R \cot\theta + 2\psi - \psi^2 \right\}(\hat{\phi}_{\mu}\hat{\theta}_{\nu} - \hat{\phi}_{\nu}\hat{\theta}_{\mu})\nonumber\\
&+& \frac{1}{r^2}\hat{\theta}^{a}\left\{ R(1-\psi)\right\}(\hat{\phi}_{\mu}\hat{\theta}_{\nu} - \hat{\phi}_{\nu}\hat{\theta}_{\mu})\nonumber\\ 
&+& \frac{1}{r^2}\left\{\hat{r}^{a}R(1-\psi)
+ \hat{\theta}^{a}(r\psi^{\prime} + R \cot\theta - R^{2})\right\}(\hat{r}_{\mu}\hat{\phi}_{\nu} - \hat{r}_{\nu}\hat{\phi}_{\mu})\nonumber\\ 
&+& \frac{1}{r^2}\hat{\phi}^{a}\left\{-(r\psi^{\prime} + \dot{R})\right\}(\hat{r}_{\mu}\hat{\theta}_{\nu} - \hat{r}_{\nu}\hat{\theta}_{\mu}),
\label{eq.10}
\end{eqnarray}
\begin{eqnarray}
D_{\mu}\Phi^{a} &=& \frac{1}{r^2}\left\{\hat{r}^{a}(r\psi^{\prime}-\psi-R^2) - \hat{\theta}^{a}R(1-\psi)\right\}\hat{r}_{\mu}\nonumber\\
&+& \frac{1}{r^2}\left\{-\hat{r}^{a}R(1-\psi) + \hat{\theta}^{a}(\dot{R}+\psi-\psi^2)\right\}\hat{\theta}_{\mu}\nonumber\\
&+& \frac{1}{r^2}\left\{\hat{\phi}^{a}(\psi-\psi^2+R\cot\theta-R^2)\right\}\hat{\phi}_{\mu}.
\label{eq.11}
\end{eqnarray}
\noindent Here prime means $\partial/\partial r$ and dot means $\partial/\partial \theta$.
The gauge fixing condition that we used here is the radiation or Coulomb gauge, $\partial^i A^a_i = 0$, $A^a_0 = 0$. 

The ansatz (\ref{eq.8}) is substituted into the equations of motion (\ref{eq.4}) as well as the Bogomol'nyi equations with the positive sign and the resulting equations of motion are just two first order differential equations,
\begin{equation}
r\psi^{\prime} + \psi - \psi^2 = -p,\\
\label{eq.12}
\end{equation}
\begin{equation}
\dot{R} + R\cot\theta - R^2 = p,
\label{eq.13}
\end{equation}

\noindent where $p$ is an arbitary constant. Eq.(\ref{eq.12}) is exactly solvable for all real values of $p$ and the integration constant can be scaled away by letting $r\rightarrow r/c$, where $c$ is the arbitrary integration constant. Hence without any loss in generality, $c$ is set to unity. In order to obtain solutions of $\psi$ with $(2m+1)$ powers of $r$ we can write $p = m(m+1)$ where $m$ is real. By doing so, the solutions of the Riccati equation (\ref{eq.13}) can be exactly solved in terms of the Legendre functions of the first and second kind. For the solutions of Eq.(\ref{eq.13}) to be regular along the z-axis, we require $R(\theta)$ to vanish when $\theta=0$ and $\theta=\pi$. To achieve these boundary conditions, the integration constant of Eq.(\ref{eq.13}) is set to zero and $m$ is restricted to take integer values. The solutions for $\psi$ and $R$ are then given respectively by
\begin{eqnarray}
\psi(r) &=& \frac{(m+1)-m r^{2m+1}}{1+r^{2m+1}},\nonumber\\
R(\theta) &=& (m+1)\left\{\cot\theta-\frac{P_{m+1}(\cos\theta)}{P_m(\cos\theta)}\csc\theta\right\}, 
\label{eq.14}
\end{eqnarray}
where $P_m$ is the Legendre polynomial of degree $m$, and $m=0, 1, 2, 3, \dots $. Hence the boundary conditions of the solutions, Eq.(\ref{eq.14}), are ~$\psi(0) = m+1, ~~\psi(\infty) = -m, ~~R(0) = R(\pi) = 0$.

In the BPS limit, the energy can be written in the form 

\begin{eqnarray}
E &=& \mp\int\partial_i(B^a_i\Phi^a)~d^3 x + \int\frac{1}{2}(B^a_i \pm D_i\Phi^a)^2~d^3 x\nonumber\\
&=& \mp\int\partial_i(B^a_i\Phi^a)~d^3 x = 4\pi M\frac{\mu}{\sqrt{\lambda}},
\label{eq.15}
\end{eqnarray}
where $M$ is the ``topological charge" when the vacuum expectation value of the Higgs field, $\frac{\mu}{\sqrt{\lambda}}$ is non zero coupled with some non-trivial topological structure of the fields at large $r$. 

The energy density $\partial_i(B^a_i\Phi^a)$ of the non-Abelian system is finite everywhere and vanishes as $1/r^2$ at large $r$ except at the origin $r=0$ due to the presence of a point singularity there and along singular planes where $P_m(\cos\theta)$ vanishes.  The energy of the system as given by Eq.(\ref{eq.15}) is finite if the integration over all space is performed only over regions of non-singularity. The two antimonopoles which are regular antimonopoles and the vortex rings are located at points and rings where the Higgs field, $r\Phi^a$, vanishes respectively. The monopole at the origin is of a different nature and is located where the Higgs field is singular. This monopole is the Wu-Yang type monopole and it possesses infinite energy density. In the Abelian gauge, this monopole carries a Dirac string singularity.

The topological charge is also related to another gauge invariant quantity of the system as given by Eq.(\ref{eq.7}),
\begin{equation}
M_\infty = \frac{1}{8\pi}\oint d^{2}\sigma_{i}\left.\left(\epsilon_{ijk}\epsilon^{abc}\hat{\Phi}^{a}\partial_{j}\hat{\Phi}^{b}\partial_{k}\hat{\Phi}^{c}\right)\right|_{r\rightarrow\infty}.
\label{eq.16}
\end{equation}
The magnetic charges enclosed by the sphere at infinity can be associated with the zeros of the Higgs field, $r\Phi^a$ and at points where $\hat{\Phi}^a$ becomes indeterminate. 
The positions of the two antimonopole do correspond to the two point zeros of the Higgs field in the A-M-A solutions. However the Wu-Yang type 1-monopole is not located at the zeros of the Higgs field but at the origin of the coordinate axes where the Higgs field is singular. 

From the ansatz (\ref{eq.8}), $A_{\mu} = \hat{\Phi}^{a}A^{a}_{\mu} = 0$. Hence from Eq.(\ref{eq.5}), the Abelian electric field is zero and the Abelian magnetic field is independent of the gauge fields $A^a_\mu$. To calculate for the 't Hooft Abelian magnetic field $B_i$, we rewrite the Higgs field of Eq.(\ref{eq.8}) from the spherical to the Cartesian coordinate system, \cite{kn:6}-\cite{kn:8} 
\begin{eqnarray}
\Phi^a &=& \Phi_{1}~\hat{r}^{a} + \Phi_{2}~\hat{\theta}^{a} + \Phi_3~\hat{\phi}^{a}\nonumber\\
&=& \tilde{\Phi}_1 ~\delta^{a1} + \tilde{\Phi}_2 ~\delta^{a2} + \tilde{\Phi}_3 ~\delta^{a3}
\label{eq.17}
\end{eqnarray}
\begin{eqnarray}
\mbox{where}~~~\tilde{\Phi}_1 &=& \sin\theta \cos \phi ~\Phi_1 + \cos\theta \cos \phi ~\Phi_2 - \sin \phi ~\Phi_3
= |\Phi|\cos\alpha \sin\beta\nonumber\\
\tilde{\Phi}_2 &=& \sin\theta \sin \phi ~\Phi_1 + \cos\theta \sin \phi ~\Phi_2 + \cos \phi ~\Phi_3
= |\Phi|\cos\alpha \cos\beta\nonumber\\
\tilde{\Phi}_3 &=& \cos\theta ~\Phi_1 - \sin\theta ~\Phi_2 = |\Phi|\sin\alpha.
\label{eq.18}
\end{eqnarray}
The Higgs unit vector is then simplified to 
\begin{eqnarray}
\hat{\Phi}^a &=& \cos\alpha \sin\beta ~\delta^{a1} + \cos\alpha \cos\beta ~\delta^{a2} + \sin\alpha ~\delta^{a3},\\
\mbox{where},~~~\sin\alpha &=& \frac{\psi\cos\theta - R \sin\theta}{\sqrt{\psi^2+R^2}},~~
\beta = \frac{\pi}{2} - \phi,\nonumber
\label{eq.19}
\end{eqnarray}
and the Abelian magnetic field is found to reduce to only the $\hat{r}_i$ and $\hat{\theta}_i$ components,
\begin{eqnarray}
B_i &=& \frac{1}{r^2 \sin\theta}\left\{\frac{\partial\sin\alpha}{\partial\theta}\frac{\partial\beta}{\partial\phi} - \frac{\partial\sin\alpha}{\partial\phi}\frac{\partial\beta}{\partial\theta}\right\}\hat{r}_i\nonumber\\
&+& \frac{1}{r\sin\theta}\left\{\frac{\partial\sin\alpha}{\partial\phi}\frac{\partial\beta}{\partial r} - \frac{\partial\sin\alpha}{\partial r}\frac{\partial\beta}{\partial\phi}\right\}\hat{\theta}_i\nonumber\\
&+& \frac{1}{r}\left\{\frac{\partial\sin\alpha}{\partial r}\frac{\partial\beta}{\partial\theta} - \frac{\partial\sin\alpha}{\partial\theta}\frac{\partial\beta}{\partial r}\right\}\hat{\phi}_i,\nonumber\\
&=& -\frac{1}{r^2 \sin\theta}\left\{\frac{\partial\sin\alpha}{\partial\theta}\right\}\hat{r}_i + 
\frac{1}{r\sin\theta}\left\{\frac{\partial\sin\alpha}{\partial r}\right\}\hat{\theta}_i.
\label{eq.20}
\end{eqnarray}
Since $\sin\alpha$ is a non singular function except at the points where the Higgs field, $r\Phi^a$ vanishes, the 't Hooft magnetic field is regular everywhere except at the locations of the A-M-A and the vortex rings. 

By noticing that the magnetic field Eq.(\ref{eq.20}) can also be written as
\begin{eqnarray}
B_i &=& \epsilon_{ijk}~\partial^j(\sin\alpha)~\partial^k\beta,\nonumber\\
&=& \epsilon_{ijk}~\partial^j(\sin\alpha~\partial^k\beta),
\label{eq.20a}
\end{eqnarray}
we found that a suitable Maxwell four-vector gauge potential for this 't Hooft magnetic field is
\begin{eqnarray}
{\cal A}_0 &=& 0,\nonumber\\
{\cal A}_i &=& (\sin\alpha - 1)\partial_i\beta = -\frac{(\sin\alpha - 1)}{r\sin\theta}\hat{\phi}_i.
\label{eq.20b}
\end{eqnarray}
This gauge potential also satisfies the Coulomb gauge condition, $\partial^i {\cal A}_i=0$. The function $\sin\alpha$ is a non singular function of $r$ and $\theta$, and it is not smooth and discontinuous only when the profile function $R(\theta)$ is singular. When $m=0$, the gauge potential Eq.(\ref{eq.20b}) is just the usual Dirac string potential and it is singular along the negative z-axis. However when $m>0$, the gauge potential ${\cal A}_i$ possesses a broken Dirac string singularity. This Dirac string singularity extends from the origin to $r=\frac{(m+1)-mr^{2m+1}}{1+r^{2m+1}}$ along the negative z-axis and from $r=\frac{(m+1)-mr^{2m+1}}{1+r^{2m+1}}$ to infinity along the positive z-axis. The gauge potential, ${\cal A}_i$, is only discontinuous at values of $\theta$ when $R(\theta)$ is singular.

From Eq.(\ref{eq.20a}), it is obvious that the magnetic field is always perpendidular to the gradient of $\sin\alpha$. Hence the magnetic field lines lie on the line $\sin\alpha = k$, $-1<k<1$, and $\phi =$ constant. By plotting $\sin\alpha = k$ on a vertical plane through the origin; we manage to draw the magnetic field lines for the configurations when $m=1$, Fig.(\ref{fig.2}); $m=2$, Fig.(\ref{fig.6}); and $m=3$, Fig.(\ref{fig.8}). 

Defining the Abelian field magnetic flux as 
\begin{eqnarray}
\Omega = 4\pi M = \oint d^{2}\sigma_{i} B_i = \int B_{i}(r^{2}\sin\theta d\theta )\hat{r}_{i}~d\phi,
\label{eq.21}
\end{eqnarray}
the magnetic charge enclosed by the sphere at infinity, $M_\infty$, is calculated to be, 
\begin{eqnarray}
M_\infty 
&=& -\left.\frac{1}{2}\sin\alpha\right|^\pi_{0, r\rightarrow \infty} = -1,~~~\mbox{when}~~m=1, 2, 3, \dots\nonumber\\
&=& 1,~~~\mbox{when}~~m=0.
\label{eq.22}
\end{eqnarray}
From Eq. (\ref{eq.22}), we can conclude that the total magnetic charge $M$ of these axially symmetric solutions does not depend on the degree of the Legendre polynomial when $m>0$. Hence the net magnetic charge of the system when $m>0$ is always negative one. By letting $M_0$ to be the net magnetic charge when the radius of the enclosing sphere tends to zero at the origin, we get,
\begin{eqnarray}
M_0 = -\left.\frac{1}{2}\sin\alpha\right|^\pi_{0, r\rightarrow 0} = 1,~~~m=0, 1, 2, 3, \dots.
\label{eq.23}
\end{eqnarray}
Similiarly, the net magnetic charge, $M_0$, at the point singularity of the solution is independent of the value of $m$. In fact, it is true that for positive non zero $m$, when $r<\sqrt[2m+1]{\frac{m+1}{m}}$, the topological magnetic charge is one, and when $r>\sqrt[2m+1]{\frac{m+1}{m}}$, the topological magnetic charge is negative one. Hence there is a 1-monopole located at $r=0$ and two antimonopoles located along the z-axis at $r=\sqrt[2m+1]{\frac{m+1}{m}}$.

We also notice that we can write the net magnetic flux per $4\pi$ sterad passing through the spherical surface of a partial enclosing sphere of radius $r$, sustaining an angle $\theta$ at the origin with the positive z-axis as
\begin{eqnarray}
M_r(\theta) &=& -\left.\frac{(\psi(r)\cos\theta-R(\theta)\sin\theta)}{2\sqrt{\psi^2(r)+R^2(\theta)}}\right|_0^\theta\nonumber\\ 
&=& \frac{1}{2}\left\{\frac{\psi}{|\psi|} - \sin\alpha\right\}.
\label{eq.24}
\end{eqnarray}

\section{Monopole, Antimonopoles, and Vortex Rings}
\label{sect:4}
The first member of this series of axially symmetric solutions is when $m=0$. As discussed in our previous work \cite{kn:7}, this solution is the Wu-Yang type monopole located at $r=0$. This radially symmetric monopole with its magnetic field, $B_i = \frac{1}{r^2}\hat{r}_i$, has the vector potential ${\cal A}_i$ given by Eq.(\ref{eq.20b}). This is just the Dirac string gauge potential, ${\cal A}_\mu = \frac{1}{r}\tan(\frac{1}{2}\theta)\hat{\phi}_\mu$, which is singular along the negative z-axis.

When $m=1$, the configuration is the second member of the axially symmetric monopole solutions. This configuration is similar to the A1 solution of Ref.\cite{kn:7} and \cite{kn:8} with gauge potentials and Higgs field given by 
\begin{eqnarray}
A_{\mu}^a &=& \frac{1}{r}\left\{\frac{2-r^3}{1+r^3}\right\}\left(\hat{\theta}^{a}\hat{\phi}_\mu - \hat{\phi}^{a}\hat{\theta}_{\mu}\right)  + \frac{1}{r}\tan\theta\left(\hat{\phi}^{a}\hat{r}_{\mu} - \hat{r}^{a}\hat{\phi}_{\mu}\right),\nonumber\\
\Phi^a &=& \frac{1}{r}\left\{\frac{2-r^3}{1+r^3}\right\}~\hat{r}^a + \frac{1}{r}\tan\theta~\hat{\theta}^a.
\label{eq.25}
\end{eqnarray}
It was first thought to consist of a 1-monopole at $r=0$, surrounded by two antimonopoles located at the point zeros of the Higgs field at $z=\pm \sqrt[3]{2}= \pm 1.2599$. However, a closer study of this solution reveals that the 1-monopole actually has a zero size \textcolor{blue}{MAM} structure along the z-axis and hence possess a net unit magnetic charge. This \textcolor{blue}{MAM} structure can be read from the plots of Eq.(\ref{eq.24}), $M_r(\theta)$ versus $\theta$ for the cases of $r\rightarrow 0$ and $r\rightarrow \infty$, see Fig.(\ref{fig.1}). The plot at $r\rightarrow \infty$ indicates that there is zero flux through the spherical shell at infinity when $\theta\not=\frac{\pi}{2}$ rad. Hence all the flux at infinity is radially inwards along the equatorial plane towards the origin, $r=0$, giving a net topological charge of negative one for the $m=1$ configuration. The radial component of the magnetic field at large $r$ is just a Dirac delta function of $\theta$, given by $B_r = -\frac{2}{r^2\sin\theta}\delta(\theta-\frac{\pi}{2})$. Hence the singularity of the solution $R(\theta)$ at the equatorial plane $\theta=\pi/2$ gives rise to a Dirac delta function singularity in the magnetic field. The antimonopole at the positive z-axis interacts with the nearest \textcolor{blue}{MAM} 1-monopole at $r=0$ to form a dipole pair and similarly the antimonopole at the negative z-axis interacts with the other \textcolor{blue}{MAM} 1-monopole at $r=0$ to form the other dipole pair leaving behind an antimonopole at the origin, see Fig.(\ref{fig.2}). At large distances all the magnetic field above and below the equatorial plane are being screened off by the two dipole pairs along the z-axis leaving behind a Dirac delta function magnetic field along the plane of singularity, $\theta=\pi/2$. 

At finite $r<\sqrt[3]{2}$, the radial component of the magnetic field is a regular function of $r$ and $\theta$ but not smooth at $\theta=\pi/2$. In fact, at $\theta=\pi/2$, $B_r$ possesses a negative Dirac delta function singularity indicating an antimonopole at the center of the composite 1-monopole. 
Fig.(\ref{fig.1}) shows that at small $r$, the net flux through the upper ($0<\theta<\frac{\pi}{2}$) and lower ($\frac{\pi}{2}<\theta<\pi$) spherical shell is $+4\pi$ each and the flux through the circle at constant $r$ and $\theta=\frac{\pi}{2}$ is $-4\pi$, hence indicating a \textcolor{blue}{MAM} structure for the 1-monopole at the origin. 

By plotting the magnetic field lines of this configuration we can confirm that at large $r$, all the magnetic field lies in the equatorial plane and is pointing radially inwards as the net magnetic charge $M_\infty$ is $-1$. A plot of the magnetic field lines is shown in Fig.(\ref{fig.2}). Hence the pole at the center of the composite monopole is an antimonopole surrounded by two 1-monopoles at zero range from each other and yet they do not annihilate each. The antimonopoles situated at $z=\pm \sqrt[3]{2}$ form dipole pairs with the nearest 1-monopoles of the \textcolor{blue}{MAM} structure, thus screening off all the magnetic field above and below the equatorial plane at $r$ infinity. There is no vortex ring in this configuration.
The Abelian gauge potential, ${\cal A}_\mu = -\frac{(\sin\alpha - 1)}{r\sin\theta}\hat{\phi}_\mu$,  possesses a Dirac string singularity along negative $z$-axis for $0<r<\sqrt[3]{2}$ and along the positive $z$-axis for $r>\sqrt[3]{2}$ to infinity.

The vortex ring appears when $m=2$, that is, when the gauge field potentials and Higgs field are respectively
\begin{eqnarray}
A_{\mu}^a &=& \frac{1}{r}\left\{\frac{3-2r^5}{1+r^5}\right\}\left(\hat{\theta}^{a}\hat{\phi}_\mu - \hat{\phi}^{a}\hat{\theta}_{\mu}\right)  + \frac{1}{r}\left\{\frac{6\cos\theta\sin\theta}{3\cos^2\theta-1}\right\}\left(\hat{\phi}^{a}\hat{r}_{\mu} - \hat{r}^{a}\hat{\phi}_{\mu}\right),\nonumber\\
\Phi^a &=& \frac{1}{r}\left\{\frac{3-2r^5}{1+r^5}\right\}~\hat{r}^a + \frac{1}{r}\left\{\frac{6\cos\theta\sin\theta}{3\cos^2\theta-1}\right\}~\hat{\theta}^a.
\label{eq.26}
\end{eqnarray}
The plots of the magnetic flux, Eq.(\ref{eq.24}), versus $\theta$, for values of $r\rightarrow 0$ and $r$ at infinity, Fig.(\ref{fig.3}), reveal that the 1-monopole at $r=0$ has a MA\textcolor{blue}{M}AM structure. However only the monopole at the center has unit charge. All the other four poles have charge less than unity. Hence there exist a vortex point both above and below the 1-monopole.

The two outer regular antimonopoles are located at the two point zeros of the Higgs field at $z=\pm \sqrt[5]{3/2} = \pm 1.0845$, Fig.(\ref{fig.4}), and the vortex ring is located along the ring of radius $1.0845$ on the equatorial plane where the Higgs field vanishes, Fig.(\ref{fig.5}). The magnetic field lines of this one vortex ring solution is shown in Fig.(\ref{fig.6}).

The two vortex rings solution is the next solution of this series of axially symmetric monopole configurations with parameter $m=3$. The gauge field potentials and Higgs field are respectively given by 
\begin{eqnarray}
A_{\mu}^a &=& \frac{1}{r}\left\{\frac{4-3r^7}{1+r^7}\right\}\left(\hat{\theta}^{a}\hat{\phi}_\mu - \hat{\phi}^{a}\hat{\theta}_{\mu}\right)  + \frac{3\tan\theta}{r}\left\{\frac{5\cos^2\theta-1}{5\cos^2\theta-3}\right\}\left(\hat{\phi}^{a}\hat{r}_{\mu} - \hat{r}^{a}\hat{\phi}_{\mu}\right),\nonumber\\
\Phi^a &=& \frac{1}{r}\left\{\frac{4-3r^7}{1+r^7}\right\}~\hat{r}^a + \frac{3\tan\theta}{r}\left\{\frac{5\cos^2\theta-1}{5\cos^2\theta-3}\right\}~\hat{\theta}^a.
\label{eq.27}
\end{eqnarray}
As usual the two point antimonopoles are situated at the two point zeros, $z=\pm \sqrt[7]{4/3} = \pm 1.0420$, of the Higgs field. The two vortex rings are located horizontally at $r=1.0420$ and $\theta=1.1071, (\pi-1.1071)$ rad. Again from the plots of magnetic flux, $M_r(\theta)|_{r\rightarrow 0}$ and $M_r(\theta)|_{r\rightarrow \infty}$, Fig.(\ref{fig.7}), of Eq.(\ref{eq.24}), together with the plot of the magnetic field lines, Fig.(\ref{fig.8}), we can conclude that the structure of the composite 1-monopole at the origin is MA\textcolor{blue}{MAM}AM, with an antimonopole at the center. The three poles in the center, \textcolor{blue}{MAM}, possess unity charge whereas the other four poles possess charge less than unity. Hence there exist a vortex point both above and below the \textcolor{blue}{MAM} 1-monopole.
Hence by induction, we conclude that the number of A and M ``poles" in the composite monopole is equal to $2m+1$, and when $m$ is even, the pole in the center of the structure is a monopole and when $m$ is odd, we have an antimonopole or a MAM 1-monopole in the center of the structure \cite{kn:6}. The number of vortex rings in the solution increases with $m$ and is equal to $(m-1)$. 

\section{Comments}
\label{sect:5}
We have obtained exact axially symmetric A-M-A configurations of the SU(2) YMH theory which are characterized by a positive integer parameter $m$. The 1-monopole which is located at the origin, $r=0$ where the Higgs field is singular, is a Wu-Yang type monopole. The two regular outer antimonopoles are located at the two point zeros of the Higgs field along the z-axis at $z=\pm\sqrt[2m+1]{(m+1)/m}$. When the parameter $m$ exceeds unity, vortex rings start to appear around the z-axis. The number of vortex rings in the solution is equal to $(m-1)$. 

Further investigations reveal that the 1-monopole at the origin possesses structure. It corresponds to a zero size composite monopole with its axis lying along the z-axis. By induction we conclude that the number of ``poles" in the composite monopole is given by $2m+1$. When $m$ is even, the center of the structure corresponds to a 1-monopole \textcolor{blue}{M} and when $m$ is odd, it corresponds to an antimonopole or a \textcolor{blue}{MAM} 1-monopole. We have analysed the solutions for the cases of $m=0, 1, 2,$ and $3$, with 1-monopole given by  \textcolor{blue}{M}, \textcolor{blue}{MAM}, MA\textcolor{blue}{M}AM, and MA\textcolor{blue}{MAM}AM, respectively. The MA and AM above and below the 1-monopole when $m=2$ and 3, can be thought of as vortex point as the magnetic charges of these ``poles" are less than unity. 

There are two types of singularities in solutions (\ref{eq.14}). The point singularity at the origin, $r=0$, gives rise to a Wu-Yang type monopole, \textcolor{blue}{M} or \textcolor{blue}{MAM}. This monopole possseses the usual Dirac string potential in the Abelian gauge when $m=0$. However when $m=1, 2, 3, \dots$, the Dirac string is broken into two parts. The string stretches from $r=0$ to $r=\frac{(m+1)-mr^{2m+1}}{1+r^{2m+1}}$ along the negative z-axis and from $r=\frac{(m+1)-mr^{2m+1}}{1+r^{2m+1}}$ to positive infinity along the positive z-axis.

The singularities in $R(\theta)$ when $P_m(\cos\theta)=0$ give rise to plane singularities. The number of singular planes in the solution is equal to $m$. Hence when $m=1$, the singular plane is the equatorial plane. The Abelian magnetic field possesses a negative Dirac delta function singularity along this plane, $B_r = -\frac{2}{r^2\sin\theta}\delta(\theta-\frac{\pi}{2})$. Similarly when $m=2$, the singular planes are $\theta = 0.9553$ and $2.1863$ rad and when $m=3$, the singular planes are $\theta = 0.6847$, 0, and 2.4569 rad. In all these solutions, the Abelian magnetic fields possess negative Dirac delta function singularity along these planes as the Abelian gauge potentials are discontinuous at these values of $\theta$.

Numerical static axially symmetric M-A-M-$\dots$ chain at finite poles separations has also been discussed in Ref.\cite{kn:6}. These numerical solutions belong to the topologically trivial sector when the total number of poles and antipoles is even and to the topological unit sector when the total number of poles and antipoles is odd. We have only managed to find odd total number of poles and antipoles in our solutions. Similar to the results of Ref.\cite{kn:6}, we have a monopole at the center of the composite 1-monopole when $m$ is even and an antimonopole in the center when $m$ is odd. Also similar is that our solutions have zero magnetic dipole moment as the number of poles in our solutions is odd.

Unlike the monopole solutions of Ref.\cite{kn:6}, our A-M-A poles here are of unit charge only. We did not manage to get monopoles and antimonopoles of charge equal to two units for our axially symmetric monopoles solutions. In fact we have not found any M-monopoles with finite separations when $|M|\geq 2$. 

We would also like to mention that for every monopoles, antimonopole, vortex rings solutions that we have discussed so far, there always exist an anti-configuration of the configurations discussed. This can be done by changing the $\phi$ winding number in the ansatz (\ref{eq.8}) from one to $-1$ and solving the Bogomol'nyi equation with the negative sign \cite{kn:8}. 

We would also like to mention that one-half topological magnetic charge monopole is obtained when the parameter $m$ is set to $-\frac{1}{2}$ in the solution, Eq.(\ref{eq.14}) \cite{kn:9}.  

\section*{Acknowledgements}
The authors would like to thank Universiti Sains Malaysia for the short term research grant (Account No: 304/PFIZIK/634039) and also the Academy of Sciences Malaysia for the Scientific Advancement Grant Allocation, SAGA, (CDC Code: P2/001).

\newpage

\section*{Figure Captions}
\listoffigures
\begin{figure}[tbh]
\vspace{4.8in}
\vskip1in
\hskip0.5in\special{bmp:M0_M_m=1.bmp x=5in y=5in}
\caption{A plot of $M_{r}(\theta)$ at small $r$ close to zero and $M_{r}(\theta)$ at $r$ infinity when $m=1$, versus $\theta$.}
\label{fig.1}
\end{figure}

\begin{figure}[tbh]
\vspace{5in}
\hskip0.5in\special{bmp:field_line_m=1_2.0.bmp x=5in y=5in}
\caption{A plot of the magnetic field lines when $m=1$ along a vertical plane through the z-axis. At large $r$, all the field lines are concentrated radially inwards along the equatorial plane. The two antimonopoles are located along the z-axis at $z= \pm 1.2599$.}
\label{fig.2}
\end{figure}

\begin{figure}[tbh]
\vspace{5in}
\hskip0.5in\special{bmp:M0_M_m=2.bmp x=5in y=5in}
\caption{A plot of $M_{r}(\theta)$ at $r$ close to zero and $M_{r}(\theta)$ at $r$ infinity, when $m=2$, versus $\theta$.}
\label{fig.3}
\end{figure}

\begin{figure}[tbh]
\vspace{5in}
\hskip0.5in\special{bmp:B_field_m=2_finite_r.bmp x=5in y=5in}
\caption{The Abelian magnetic field of the $m=2$ solution at finite distances showing the presence of the two dipole pairs along the z-axis and the vortex ring at $z=0, ~\rho=1.0845$.}
\label{fig.4}
\end{figure}

\begin{figure}[tbh]
\vspace{5in}
\hskip0.5in\special{bmp:B_field_m=2_vortex_ring.bmp x=5in y=5in}
\caption{The Abelian magnetic field of the $m=2$ solution at distances close to $z=0$ and $\rho=1.0845$ showing the presence of the vortex ring.}
\label{fig.5}
\end{figure}

\begin{figure}[tbh]
\vspace{5in}
\hskip0.5in\special{bmp:field_line_m=2_1.5.bmp x=5in y=5in}
\caption{A plot of the magnetic field lines when $m=2$ along a vertical plane through the z-axis. The vortex ring is situated horizontally at equal distances from the origin as the two antimonopoles at $z=\pm 1.0845$.}
\label{fig.6}
\end{figure}

\begin{figure}[tbh]
\vspace{5in}
\hskip0.5in\special{bmp:M0_M_m=3.bmp x=5in y=5in}
\caption{A plot of $M_{r}(\theta)$ at $r$ close to zero and $M_{r}(\theta)$ at $r$ infinity, when $m=3$, versus $\theta$.}
\label{fig.7}
\end{figure}

\begin{figure}[tbh]
\vspace{5in}
\hskip0.5in\special{bmp:field_line_m=3_1.5.bmp x=5in y=5in}
\caption{A plot of the magnetic field lines when $m=3$ along a vertical plane through the z-axis. The two vortex rings are situated horizontally at equal distances from the origin as the two antimonopoles at $z=\pm 1.0420$.}
\label{fig.8}
\end{figure}

\end{document}